\documentclass[12pt,twoside]{article}
\usepackage[a4paper, top=2.5cm, bottom=3.5cm, outer=2.5cm, inner=2.5cm, headsep=15pt]{geometry}
\usepackage{amsmath}
\usepackage{textcomp}
\usepackage{graphicx}
\usepackage{txfonts}
\usepackage{subfigure}
\usepackage{epstopdf}
\usepackage{natbib}
\usepackage[backref]{hyperref}
\usepackage{url}
\usepackage[T1]{fontenc}
\usepackage{csquotes}
\usepackage{fancyhdr}
\title{Forecasting of the Thermosphere via Assimilation of Electron Density and Temperature Data}
\date{\vspace{-1ex}}
\author{Timothy Kodikara\textsuperscript{1,*}~and~Kefei Zhang\textsuperscript{1,2,*}}
\begin{document}
\maketitle
\begin{flushleft}
\textsuperscript{1}~SPACE Research Centre, RMIT University, Melbourne, VIC, Australia\\
\textsuperscript{2}~School of Environment and Spatial Informatics, China University of Mining and Technology, Xuzhou 221116, P.R.~China\\
\textsuperscript{*}~email: \href{mailto:timothy.kodikara@rmit.edu.au; kefei.zhang@rmit.edu.au}{timothy.kodikara@rmit.edu.au; kefei.zhang@rmit.edu.au}
\end{flushleft}
\section*{Abstract}
The paper presents experiments of driving a physics-based thermosphere model by assimilating electron density ($Ne$) and temperature ($Tn$) data using the ensemble adjustment Kalman filter (EAKF) technique. This study not only helps to gauge the accuracy of the assimilation, to explain the inherent model bias, and to understand the limitations of the framework, but it also establishes EAKF as a viable technique in the presence of realistic data assimilation scenarios to forecast the highly dynamical thermosphere.

The results from perfect model scenarios show that data assimilation changes and, more often than not, improves the model state. Data from Swarm-A, Swarm-C, CHAMP, and GRACE-A are used to validate the resulting analysis states. The independent validation results show that the $Ne$-guided thermosphere state does not outperform the model state without data assimilation along the considered orbits. This may be due to the limited number of bonafide $Ne$ profiles available for the thermosphere specification tasks in the experiments. More importantly, the results show that the $Ne$-guided thermosphere state does not deteriorate much in performance during geomagnetic storm time. The results reveal a few challenges of using $Ne$ profiles in a hypothetical operational data assimilation exercise. In terms of estimating the mass density along the orbits of both CHAMP and GRACE-A satellites, the experiment with assimilating $Tn$ shows more promise over $Ne$.

The results show that the improvement gained in the overall forecasted thermosphere state is better during solar minimum compared to that of solar maximum. These results also provide insights into the biases inherent in the physics-based model. The systematic biases that the paper highlight could be an indication that the specification of plasma-neutral interactions in the model needs further adjustments.
\vspace{1cm}

\textbf{Key words:} space weather -- data assimilation -- Kalman filter -- forecasting -- thermosphere
\clearpage

\section{Introduction}{\label{intro_sec}}
The economy of the space industry and the safety of space assets depend, it could be argued, on our ability to model and predict space weather and the variability of the space environment. The low Earth orbit (LEO) region from about 160 to 2,000~km is where more than half of operational satellites are currently placed and also the most populated with debris \citep{Klinkrad_2017_debris}. Therefore, the risk of satellite collisions is highest in LEO and by extension the economic and social impact to the space domain is also the highest in this region \citep{MoribaJah_2018,SpaceWeatherBoard2009}. One of the largest and persistent source of uncertainty in the satellite orbit determination/prediction solutions in LEO arise from the uncertainty in atmospheric mass density estimates \citep{Vallado2014}. The uncertainty in mass density estimates is linked to imperfections in modelling this highly dynamical space environment. An accurate forecast of the mass densities is critical to precise and robust orbit predictions, as well as to assure, among others, the safety of space assets vital to many technologies on Earth \citep[e.g.][]{Hejduk_2018}.

More studies on data-guided forecasting of the thermosphere are conducted during the past decade than any other time. Yet the major thematic issues with forecasting the thermosphere remain a scientific, computing and resource challenge. These challenges include, for example, scientific---the problem of physics of the thermosphere-ionosphere system being not fully understood; computing---the problem of developing computationally efficient, operationally viable, and high-accuracy output feedback (nowcast/forecast) assimilation algorithms; and resource---the problem of scarcity of impactful measurements of the system. Some complications of building a model that includes all the relevant physics are due to the highly dynamical nature of the thermosphere that is not only driven by external forces but also controlled significantly by internal chemistry and dynamics \citep[e.g.][]{Bauer_2004}. 

Data-guided forecasting refers to the process of computing the best possible estimate of the state of the system using data along with a numerical prediction of the model state. \citet{LeeMatsuo_2012_cosmic} and \citet{Matsuo_EnKF_2013} demonstrated early success in affecting the model state in thermosphere-ionosphere-electrodynamics general circulation model (TIE-GCM; \citet{Tiegcm1981_Dickinson, Tiegcm1992Richmond}) through an ensemble Kalman filter-based (EnKF) data assimilation technique \citep[see]{Evensen_KF1994}.  The EnKF, compared to the traditional Kalman filter (KF) allows one to bypass the restrictions and limitations that are usually associated with a large complex nonlinear model such as TIE-GCM. The EnKF also affords significant computational efficiency for large geophysical models by representing model error covariance through a sample covariance computed from a series (\textit{ensemble}) of model runs. This is due to the fact that the size of the model error covariance matrix in EnKF depends on the size of the ensemble and not the size of the model dimensions (e.g. grid space, variables). Said differently, the model error depends on the degree of ensemble spread. A number of other variants of KF-based assimilation techniques and other inductive/deductive techniques, such as three-/four-dimensional variational analysis and Optimal Interpolation exist \citep[see][]{daley1993atmospheric,kalnay_2002} but these are not discussed here.

Recently, \citet{Mehta_Hermitian_2018} proposed a reduced order model (ROM) data assimilation framework for thermospheric mass density based on the proper orthogonal decomposition dimensionality reduction technique using TIE-GCM as the base model. \citet{Sutton_tiegcm_assim} proposed several changes to the EnKF/TIE-GCM assimilation framework by adopting variational techniques where the external drivers, $F_{10.7}$ and $Kp$, are estimated iteratively, and the model is iteratively run until data-model convergence is achieved with these newly estimated driver parameters. While the results in these studies show promise, the results also convey that applying these stochastic methods to the thermosphere is not necessarily simplistic. A standard EnKF/TIE-GCM framework requires about 60 ensemble members \citep[e.g.][]{Matsuo_EnKF_2013} to be processed in parallel (typically processed in multi-node supercomputers). The variational methods have a disadvantage with the nontrivial requirements of computing tangent linear models and adjoint models for both the evolution (forward) and observation operators, and ROMs usually require training with data that span over long periods (e.g. sunspot cycle).

The main goal of this work is to investigate the impact to the forecasted thermosphere state through ensemble adjustment Kalman filter (EAKF; \citeauthor{Anderson_EnKF_2001}, \citeyear{Anderson_EnKF_2001})-based assimilation of observed electron density and empirically-derived temperature into TIE-GCM. Electron density data are more globally abundant than other under-observed thermospheric parameters such as winds and mass density \citep{Matsuo_EnKF_2013} and hence they offer the most promising means to test the effect of assimilation on the model forecasted state on a global scale. Investigating the potential and limitations of assimilating temperature to assist thermosphere forecasts is useful given that the thermosphere is primarily driven by external heat and momentum sources \citep[e.g.][]{FullerRowell_1997,Kodikara_Sync,VollandHans_1988}. This work is presented as an extension to \citet{LeeMatsuo_2012_cosmic, Matsuo_EnKF_2013} and \citet{Hsu2014}. A study of the EAKF/TIE-GCM framework for the thermosphere not only helps to gauge the accuracy of the assimilation, to explain the inherent model bias, and to understand the limitations of the framework, but also establishes EAKF as a viable technique in the presence of realistic data assimilation scenarios to forecast the highly dynamical thermosphere. Sparsity of data and data uncertainty are two caveats which concern the interpretation of the assimilation results. Calibrating instrument error for real-time observations is not optimal in most cases and even non-existent for some useful satellite observations such as temperature, electron density and accelerometer-derived mass density. Therefore, accounting for observation uncertainty is a challenging task. This study employs and justifies artificially inflated data uncertainty.
\section{Data and Models}{\label{sec_models}}
The EAKF-based assimilation algorithm described in \cite{Anderson_EnKF_2001} and implemented in the Data Assimilation Research Testbed-\textit{classic} (DART; \citeauthor{Anderson_DART_2009}, \citeyear{Anderson_DART_2009}) is used to assimilate the data into TIE-GCM. The specifics of the configuration and parameter settings in DART are described in the next section.

TIE-GCM is a well-established, physics-based, self-consistent model of the thermosphere-ionosphere system that uses a finite differencing technique to discretise the numerical solutions for conservation of mass, energy and momentum \citep[e.g.][]{Maute2017,Qian201473}. Instead of the default TIE-GCM version 1.95 available in DART, the latest TIE-GCM version 2.0 (released on 21 March 2016) with a model-time-step of 30~s has been integrated into DART to perform the experiments presented here. The reader interested in more details about the open-source TIE-GCM is referred to the website at <\url{http://www.hao.ucar.edu/modeling/tgcm/tiegcm2.0}>.

In this study, the TIE-GCM outputs are recorded at a 5{\textdegree}$\times$5{\textdegree} horizontal (latitude and longitude) grid at 29 constant pressure surface layers that extend from approximately 97 to 600~km in altitude (depending on solar activity). These isobaric pressure surface layers have a resolution of half a scale-height (the scale-height in a hydrostatic atmosphere is the altitudinal difference as a result of change in air density and pressure by a factor of 1/$e$, where $e$ is the Euler's number $\sim$2.71828). In the model runs presented here, the EUVAC (extreme ultraviolet flux model for aeronomic calculations) empirical solar proxy model \citep[see][]{Richards1994,Solomon2005} is used to calculate the input from solar irradiance specified via the average of daily $F_{10.7}$ and its running 81-day centred mean $\overline{F}_{10.7}$. The high latitude mean-energy, energy flux and electric potential are specified by the \citet{Heelis1982} ion convection model, and the \citet{Roble1987} auroral particle precipitation scheme. The hemisphere power and cross-polar-cap potential drop required to determine this high latitude energy and momentum input are estimated using the $Kp$ index. In TIE-GCM the in-built wind dynamo calculates the electric potential for low and middle latitudes \citep[see][]{Tiegcm1992Richmond,Richmond1995}. The tidal forcing from the lower atmosphere is specified at the lower boundary of TIE-GCM through numerically derived migrating diurnal and semidiurnal tides using the \citet{Hagan_GSWM00_2001} global scale wave model (GSWM). In addition, day-of-year dependent perturbations to the advective and diffusive transport are introduced using the eddy diffusion coefficient as described in \citet{Qian2009}.

The electron densities used in this work are from the joint USA-Taiwan Constellation Observing System for Meteorology, Ionosphere and Climate/Formosa Satellite 3 (COSMIC/FORMOSAT-3; hereinafter COSMIC) mission. These COSMIC electron densities (hereinafter COSMIC-$Ne$) are derived from radio occultation (RO) events across the COSMIC constellation. The number of successful events is proportional to, among others, the number of GPS signal transmitters. The derivation of electron density from COSMIC RO data is detailed in \citet{Tsai2001}. The errors in the retrieval method of COSMIC-$Ne$ is widely discussed \citep[e.g.][]{Liu2009,Yue_AbleInv_2010} and many studies report a root-mean-square error (RMSE) between 10 and 20\% compared to ground measurements of electron density \citep[e.g.][]{Pedatella_2015,Yue_cosmic_2014}.

The data assimilation results are validated with independent observations from four different satellites: Swarm-A, Swarm-C, CHAllenging Minisatellite Payload (CHAMP), and Gravity Recovery and Climate Experiment (GRACE)-A. Swarm, launched in late 2013, consists of three polar (angle of inclination: 87.4{\textdegree} [A and C]; 87.8{\textdegree} [B]) orbiting satellites with an orbital period of approximately 94 minutes. The electron density from Swarm-A used here is the extended \textit{L1b} product of the Langmuir probe data from the EFI (electric field instrument). \citet{Lomidze_SwarmA_2018} report that  plasma frequency measured by Swarm is about 10\% less compared to ISR, ionosondes, and COSMIC data. The accelerometer-derived mass density $\rho$ from Swarm-C used here is the postprocessed \textit{Level2daily} (SW\textunderscore{OPER}\textunderscore{DNSCWND}\textunderscore{2}) product \citep{Siemes2016}. The methods applied in Swarm-C data to isolate the drag-acceleration and derive the mass density are described in \citet{Doornbos_2012}. \citet{Kodikara_Swarm} provides a comparison between Swarm-C accelerometer-derived thermospheric mass density and multiple model estimates including the TIE-GCM. 

The $\rho$ data from CHAMP and GRACE-A satellites that have been recalibrated by \citet{Mehta_2017} using physics-based drag coefficients are used here to validate the data assimilation results. \citet{Mehta_2017} report an average bias of 14--18\% for CHAMP and 10--24\% for both GRACE-A and -B with respect to mass densities in both \citet{Sutton_2011champgrace} and \citet{Doornbos_2012}. The orbital periods of CHAMP and GRACE-A for the experiments considered here are 91 and 94~min, respectively. This work also explores the impact of assimilating temperature $Tn$ along the CHAMP orbit (hereinafter CHAMP-$Tn$) on the global thermospheric mass density state. The empirically-derived $Tn$ data used here are from \citet{Mehta_2017}'s above-mentioned recalibrated data set.
\section{Ensemble Adjustment Kalman Filter Experiments}{\label{sec_assim}}
\begin{figure}
\centering
\includegraphics[width=0.7\columnwidth]{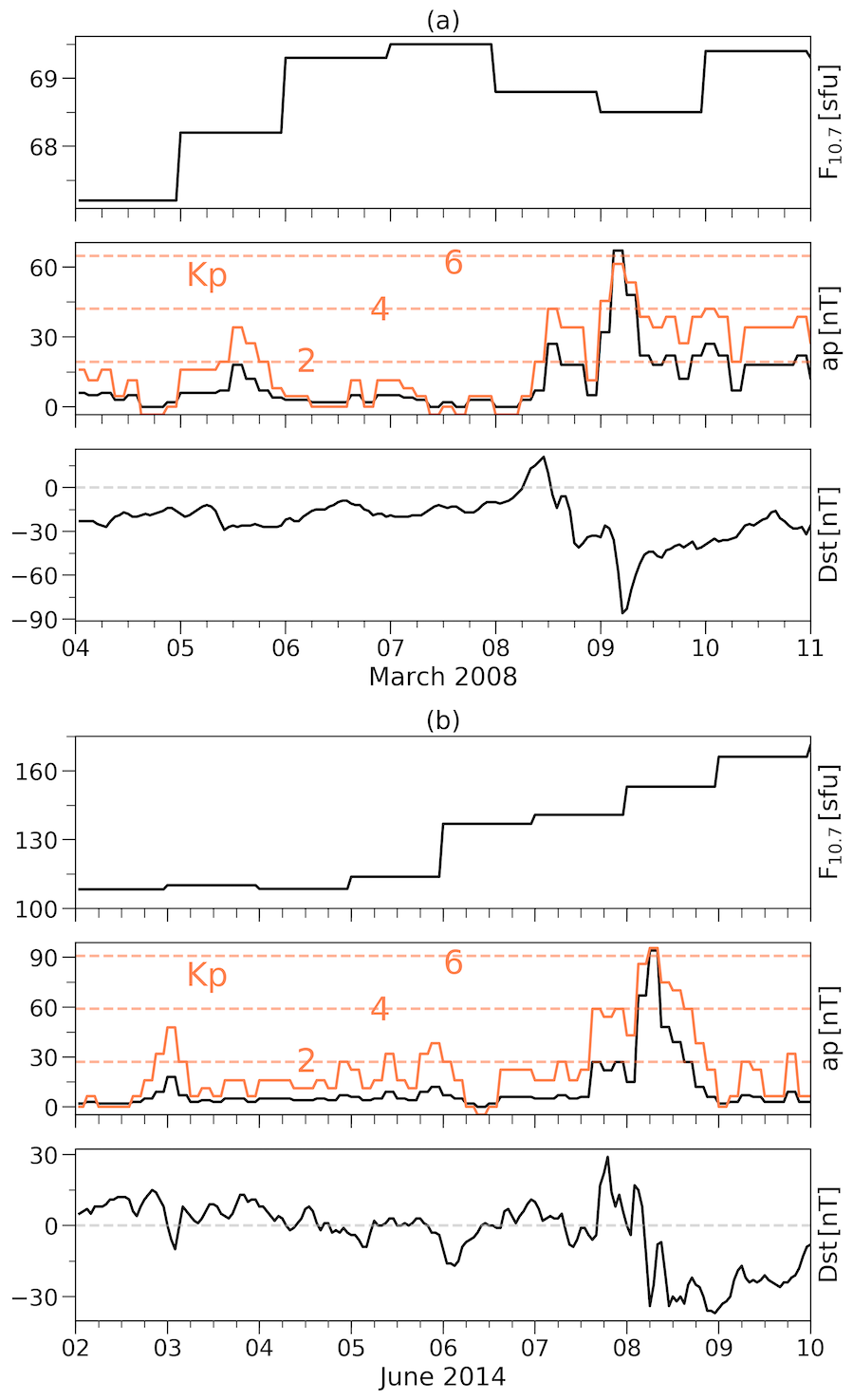}
\caption{Space weather conditions for (a) March 2008 (E1, E3) and (b) June 2014 (E2) demonstrated via $F_{10.7}$ solar flux, $ap$, and the $Dst$ indices. The corresponding $Kp$ values are overlaid on $ap$ and marked in orange. Source: OMNI data available on omniweb.gsfc.nasa.gov.\label{solgeoMar_Jun}}
\end{figure}
Briefly, the EAKF---similar to the EnKF described in \citet{Evensen_KF1994}, is a Monte-Carlo approximation of the more traditional KF \citep{Kalman1960,Anderson_EnKF_2001}. It provides a relatively easy mechanism to assimilate observations into TIE-GCM to estimate the impact of the observations on the model state forward in time. Similar to the EnKF, EAKF uses sample statistics (means and covariances) from the prior ensemble of model states to calculate the posterior probability distributions. The probability distribution prior to the assimilation of data is referred to as the prior. The posterior is the probability distribution of the prior distribution updated with observations. Unlike the EnKF, the EAKF does not add noise from a sample of perturbed observations but applies the linear operator described in \citet{Anderson_EnKF_2001} to update the prior ensemble of model states that yield theoretically consistent means and covariances. 

The following three experiments are presented here to analyse the ability of the EAKF technique to guide TIE-GCM:
\begin{enumerate}
\item[E1] Assimilate COSMIC-$Ne$ during solar minimum (2008 March 4--11);
\item[E2] Assimilate COSMIC-$Ne$ during solar maximum (2014 June 2--10); and
\item[E3] Assimilate CHAMP-$Tn$ during solar minimum (2008 March 4--11).
\end{enumerate}

Assimilation of data during each period starts at 1~UT and ends at 0~UT on the respective days.
The periods in 2008 and 2014 are selected to test the effectiveness of the assimilation during solar minimum and solar maximum, respectively. The other criterion applied in selecting these periods is that they be relatively quiet in geomagnetic activity with at least one geomagnetic storm event.

Figure~\ref{solgeoMar_Jun}a (\ref{solgeoMar_Jun}b) presents the $F_{10.7}$ solar flux, $ap$, $Kp$, and $Dst$ indices to illustrate the space weather conditions during the period corresponding to E1 and E3 (E2). The geomagnetic activity indicated by $ap$ and $Dst$ during the first few days of both assimilation periods is largely quiet. Two relatively strong geomagnetic storms occur on 9 March 2008 and 8 June 2014. While both of these events that lie in the $ap$ range 65--90~nT share a daily average of approximately 6 in the $Kp$ index, the storm on 9 March 2008 is stronger in terms of the $Dst$ index, which indicates that the two events are significantly different from each other. $Dst$ is an index that indicates the disturbance on the equatorial magnetospheric ring-current.The solar activity during June 2014 is higher than that of March 2008 and increases from approximately 105 to 171~sfu. The geophysical indices $F_{10.7}$ solar flux and $Kp$ (hereinafter GPI) shown in Figure~\ref{solgeoMar_Jun} are used to drive the TIE-GCM to obtain the corresponding control state $\boldsymbol{x}^{c}$ for E1, E2, and E3. The two periods provide an opportunity to compare the model error growth in the EAKF at vastly different solar activity levels but seemingly similar geomagnetic conditions with different storm characteristics.

The EAKF state vector $\boldsymbol{x}$ for these experiments is selected in such a way to avoid/minimise spurious strong correlations between observations and model variables (e.g. winds, temperature, and composition). In this EAKF/TIE-GCM framework, the observation operators in DART compute the expected value of an observation given the model state. The size of the observation vector in the above-mentioned experiments (typical in atmospheric data assimilation) is much smaller than the size of the state vector. In a nutshell, spurious correlations occur as a result of long spatial distances between observation and model variable---spatially remote observations, and due to certain model variables being physically unrelated to the observation \citep{Anderson_EnKF_2001}.

Specific localisation methods are applied to combat this problem of spurious correlations. The \citet{Gaspari_1999} correlation function is used to constrain the spatial region of the impact of the observation. In the experiments presented here, the correlation function shifts the impact of the observation from a maximum at the observation location to zero at the cutoff distance following an approximation of a Gaussian curve. Spurious correlations between observations and physically unrelated model variables may be generated from limitations of the ensemble size, which is also much less than the size of the state vector \citep{Anderson_EnKF_2001}. A subset of TIE-GCM prognostic variables/fields that are known to be strongly correlated with mutual physical relationships is selected as the EAKF state vector to be updated in each assimilation cycle and in turn affect the model forecast. \citet{LeeMatsuo_2012_cosmic}, \citet{Matsuo_EnKF_2013}, \citet{Hsu2014}, and \citet{Chartier2016} presented results from different combinations of prognostic variables in the state vector. Overall, they demonstrated that inferring the dynamical state of both ionosphere and thermosphere is improved by including thermosphere-ionosphere coupling parameters such as electron density, temperature, winds and composition in the state vector. The EAKF state vector $\boldsymbol{x}$ selected for this study is analogous to the superior-performing state vector in \citet{Hsu2014}.

The EAKF state vector $\boldsymbol{x}$ in the experiments mentioned above is composed of,
\[
\boldsymbol{x}  = [\boldsymbol{\psi}^{Tn}; \boldsymbol{\psi}^{\gamma \textrm O}; \boldsymbol{\psi}^{\gamma \textrm O^{+}}; \boldsymbol{\psi}^{\gamma \textrm O_{2}}; \boldsymbol{\psi}^{U} ; \boldsymbol{\psi}^{V}; \boldsymbol{\psi}^{Ne}],
\]
where ${Tn, \textrm O, \textrm O^{+}, \textrm O_{2}, U, V}$, and $Ne$ represent temperature [K], atomic oxygen [$\gamma$], atomic oxygen ion [$\gamma$], molecular oxygen [$\gamma$], zonal (east-west) wind [m$\cdot$s$^{-1}$], meridional (north-south) wind [m$\cdot$s$^{-1}$], and electron number density [cm$^{-3}$], respectively. $\psi$ denotes the full vector of the respective prognostic variable over the entire model space. Although the size of the assimilating observation vector may change in size with time, the size of $\boldsymbol{x}$ is constant. The mass mixing ratio $\gamma$ of the major species is obtained with the assumption $\gamma\textrm N_{2}~=~1 - \gamma\textrm O - \gamma\textrm O_{2}$, thus affecting also the $\gamma\textrm O/\gamma\textrm N_{2}$ in $\boldsymbol{x}$. At the upper boundary, $\gamma \textrm O$ and $\gamma \textrm O_{2}$ are considered to be in diffusive equilibrium. At the lower boundary, the vertical gradient of $\textrm O$ is taken to be zero and $\gamma \textrm O_{2}$ is fixed at 0.22.

The thermosphere is driven by external heat and momentum sources, which are primarily characterised in TIE-GCM by the GPIs. The model error growth in EAKF is represented by the degree of spread among the ensemble of model states. If the GPIs are held constant then the probability distribution represented by the ensemble have no means of characterising the effects of driver uncertainty on the model error growth \citep{Matsuo_EnKF_2013}. Therefore, to aid the characterisation of model error growth, the ensembles for all three experiments are generated by perturbing the primary forcing parameter $\boldsymbol{d}$ for each ensemble member $m$, where
\[
\boldsymbol{d}^{(m)}~=~[F_{10.7}^{(m)},~\overline{F}_{10.7}^{(m)},~Kp^{(m)}].
\]
$F_{10.7}, \overline{F}_{10.7}$ and $Kp$ for $\boldsymbol{d}$ are sampled from a normal distribution as follows:
\[
\boldsymbol{d}^{(m)} \hookleftarrow  \mathcal{N}\Big(\big[\mu_{F_{10.7}}, \mu_{\overline{F}_{10.7}}, \mu_{Kp}\big],~\big[\sigma_{F_{10.7}}^{2}, \sigma_{\overline{F}_{10.7}}^{2}, \sigma_{Kp}^{2}\big]\Big),
\]
with $\{Kp \mid Kp\ge1\}$. The mean $\mu$ values for $F_{10.7}, \overline{F}_{10.7}$, and $Kp$ for the experiments in March 2008 are 68~sfu, 60~sfu, and 3, respectively. The corresponding $\mu$ values for the experiment in June 2014 are 140~sfu, 130~sfu, and 4. Their respective $\sigma^{2}$ values give the variance of the perturbation for $F_{10.7}, \overline{F}_{10.7}$, and $Kp$. The $\sigma_{F_{10.7}}, \sigma_{\overline{F}_{10.7}}$, and $\sigma_{Kp}$ for the experiments in March 2008 are 15~sfu, 15~sfu, and 2, respectively. And the corresponding $\sigma$ values for the experiment in June 2014 are 30~sfu, 30~sfu, and 2. The widths of these $\boldsymbol{d}$ distributions are thus specified considering the background GPIs shown in Figure~\ref{solgeoMar_Jun}. Each ensemble member is randomly assigned a combination of forcing parameters that are kept fixed throughout the entire assimilation.

The specifics of the DART configuration and parameter settings used here are as follows:
\begin{enumerate}
\item The ensemble size is 60 for each experiment;
\item The model error covariance is localised using the \citet{Gaspari_1999} correlation function with a half-width of 0.2 radians horizontally;
\item The vertical localisation height is set equal to 200~km for E1 and E2, and 40~km for E3;
\item The outlier threshold for observations is three standard deviations from the prior ensemble mean;
\item The assimilation window is 3,600~s for E1 and E2, and 5,400~s for E3---centred at current model-time;
\item Spatially-varying state space inflation is applied to the prior state before observations are assimilated with initial inflation, inflation standard deviation, and inflation damping set equal to 1.02, 0.6, and 0.9, respectively;
\item The minimum $Ne$ is 1,000~cm$^{-3}$;
\item The lower bound of the temperature is 100~K; and
\item The $\gamma\textrm O$ and $\gamma\textrm O_{2}$ have cutoff limits at zero and one.
\end{enumerate}
A brief commentary on the above-mentioned parameter settings:

\citet{Matsuo_EnKF_2013} report no significant difference in the quality of the analysis state for ensemble sizes above 60.  However, their experiments are conducted with 90- to 100-member ensembles. The half-width for the \citet{Gaspari_1999} correlation function used in this study is similar to \citet{Matsuo_EnKF_2013} and \citet{Hsu2014}. \citet{Chartier2016} considering the effect on total electron content over the continental USA showed no appreciable difference between the use of localisation half-width radii 0.2, 0.5, and 1.0 radians under geomagnetically quiet times. In order to smoothen the effect of the assimilation on the vertical profile of model electron density, \citet{Matsuo_EnKF_2013} also used a vertical localisation height of 200~km for the experiment with COSMIC-$Ne$. A smaller vertical localisation height is used for E3 as the assimilating data are also essentially localised along the CHAMP orbit---unlike COSMIC-$Ne$ data, which have a larger 3D coverage than CHAMP-$Tn$. In other words, a large vertical localisation height for E3 is prone to spurious correlations.

The COSMIC-$Ne$ values less than zero and outside the above-mentioned outlier threshold are removed from the experiments. \citet{Chartier2016} employed a similar threshold to control the impact of observations. The total number of COSMIC-$Ne$ profiles ingested into the observational forward operator is 54501 and 61919 for E1 and E2, respectively. Similarly, 60479 CHAMP-$Tn$ epochs are used in E3. The assimilation windows applied in this study is similar to \citet{Matsuo_EnKF_2013}. A 90-min assimilation window is used for E3 to correspond with the approximate orbital period of CHAMP. The above-mentioned spatially-varying state space inflation values are thus specified on an ad-hoc basis following the recommendation in DART documentation for large geophysical models. It is unnatural to have values below the above-mentioned minimum $Ne$ and minimum temperature in the altitude ranges considered in the study. \citet{LeeMatsuo_2012_cosmic} also employed similar parameter settings for $Ne$, $\gamma\textrm O$, and $\gamma\textrm O_{2}$.

The success of the experiments is assessed using independent satellite data. In addition, the effectiveness of the EAKF technique independent of model bias is assessed in a ``perfect model'' scenario. The control state $\boldsymbol{x}^{c}$ (truth) for the perfect model scenario for all three experiments is a TIE-GCM run each driven by observed GPIs. All model runs including the ensemble members are primed with a ``spin-up'' period of 15 days prior to the assimilation. The mean of the updated/posterior ensemble is referred to as the analysis state $\boldsymbol{x}^{a}$ and the mean of the prior ensemble is referred to as the forecast state $\boldsymbol{x}^{f}$. In the results presented below, the DART assimilation results are assessed using these analysis and forecast states.

In the perfect model scenario, as per \citet{Hsu2014}, the ratio of RMSE between $\boldsymbol{x}^{a}$ and $\boldsymbol{x}^{f}$ each with respect to $\boldsymbol{x}^{c}$ is used to evaluate the impact to the model state with and without data assimilation. The ratio of RMSE $R_{\textrm rmse}$ is computed from,
\begin{equation}
R_{\textrm rmse}(\psi)=\frac{\sqrt{\big(\boldsymbol{x}^{a}(\psi) - \boldsymbol{x}^{c}(\psi)\big)^{2}}}{\sqrt{\big(\boldsymbol{x}^{f}(\psi) - \boldsymbol{x}^{c}(\psi)\big)^{2}}},
\label{eqn_Rrmse}
\end{equation}
where $\{R_{\textrm rmse}\in\mathbb R~:~R_{\textrm rmse}\ge 0\}$ and $\psi$ denotes the prognostic/diagnostic variable in TIE-GCM. 
\begin{figure}
\centering
\includegraphics{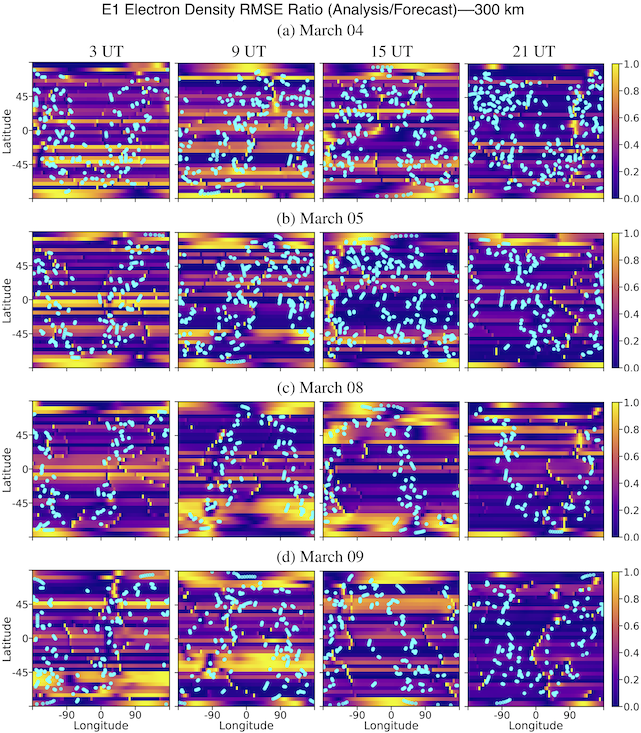}
\caption{(a--d) The geographic latitude-longitude distribution of the RMSE ratios of electron density ($R_{\textrm rmse}$($Ne$)) for E1 at 300-km altitude. The blue dots indicate the locations of assimilated COSMIC-$Ne$ profiles in the 100- to 500-km altitude range and within $-$2.5 and $+$0.5 hr of a given UT. The $R_{\textrm rmse}$($Ne$) are scaled from 0 to 1 where values close to 0 indicate that the analysis state $\boldsymbol{x}^{a}$ is closer to the ``truth''.\label{MarNEratio300}}
\end{figure}

\begin{figure}
\centering
\includegraphics{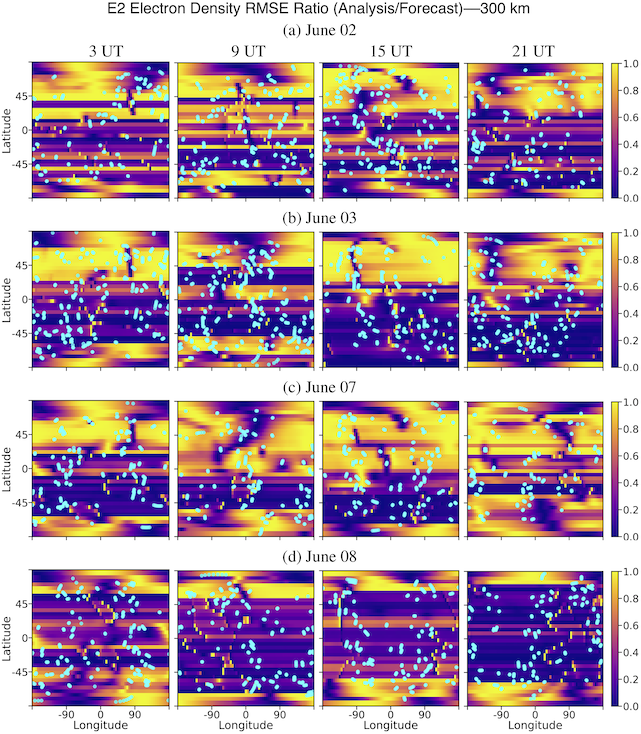}
\caption{Same as Figure~\ref{MarNEratio300} except for E2.\label{JunNEratio300}}
\end{figure}
\section{Results}{\label{results}}
\subsection{Impact of Assimilating COSMIC-$Ne$ Profiles}\label{sec_Neassim}
Figures~\ref{MarNEratio300} and \ref{JunNEratio300} compare the $R_{\textrm rmse}$($Ne$) at 300-km altitude for E1 and E2, respectively. The two figures provide a digest of the impact of the assimilation towards the beginning of the assimilation (4--5 March 2008; 2--3 June 2014) and the end of the assimilation (8--9 March 2008; 7--8 June 2014) over four different UTs. In Figures~\ref{MarNEratio300} and \ref{JunNEratio300}, the $R_{\textrm rmse}$($Ne$) is scaled from 0 to 1, where 0 indicates that the analysis $\boldsymbol{x}^{a}$($Ne$) is statistically indistinguishable from the control state $\boldsymbol{x}^{c}$($Ne$)---the ``truth''. The non-scaled minimum and maximum of the $R_{\textrm rmse}$($Ne$) spread is similar between the two figures and the range of the spread is relatively small globally. Therefore, a unified scaling highlights the areas where $\boldsymbol{x}^{a}$ is impacted from the assimilation across the different UTs better than the non-scaled $R_{\textrm rmse}$. The blue dots represent the locations of the assimilated COSMIC-$Ne$ profiles in the 100- to 500-km altitude range and within $-$2.5 and $+$0.5 hr of a given UT. The geographic latitude-longitude resolution of the two figures is the same as the model grid resolution, which is 5{\textdegree}$\times$5{\textdegree}.

In general, the impact of the assimilation is less in more areas in E2 than E1 as indicated by the yellow areas in Figures~\ref{MarNEratio300} and \ref{JunNEratio300}. A relationship between the assimilated COSMIC-$Ne$ profiles (blue dots) and the $R_{\textrm rmse}$ results is apparent in both figures. For example, the number of assimilated observations in Figure~\ref{MarNEratio300}a [3~UT] is less compared to that of in Figure~\ref{MarNEratio300}a [21~UT] where the assimilated state is also more improved globally (purple areas) than at 3~UT.  The $R_{\textrm rmse}$ is larger in the high southern latitudes between 180{\textdegree}W and 0{\textdegree} longitude in Figure~\ref{MarNEratio300}c [15~UT] where there are also not many observations. On the contrary, in a few areas with clusters of observations, a higher $R_{\textrm rmse}$ is also present (e.g. below 45{\textdegree}S and between 90{\textdegree}E and 180{\textdegree}E in Figure~\ref{MarNEratio300}c [9~UT], and below 45{\textdegree}S and between 180{\textdegree}W and 90{\textdegree}W in Figure~\ref{MarNEratio300}d [3~UT]).

The relationship between observation locations and $R_{\textrm rmse}$-improved areas is more distinct in Figure~\ref{JunNEratio300} than in Figure~\ref{MarNEratio300}. In Figure~\ref{JunNEratio300}, the improved areas are mostly concentrated in the Southern Hemisphere where the density of assimilated observations is higher compared to the Northern Hemisphere. Interestingly, for example, Figure~\ref{JunNEratio300}d [21~UT] shows an improvement in the northern latitudes (up to approximately 70{\textdegree}N) where more observations are present in the Northern Hemisphere compared to the previous day. Overall, both Figures~\ref{MarNEratio300} and \ref{JunNEratio300} show that assimilating COSMIC-$Ne$ has changed the model state and in more areas than not it has reduced the $R_{\textrm rmse}$. The distinct areas where the relationship between the assimilated COSMIC-$Ne$ profiles and the $R_{\textrm rmse}$ results is not clear or seems to be inverse highlights the complexity of the thermospheric dynamics. In other words, the correlation between 3~hr of observations, and the evolving differences between the analysis and forecast states itself is intricate as the snapshots shown here may not capture the impact of the assimilated observations in an accommodating time-scale.
\begin{figure}
\centering
\includegraphics{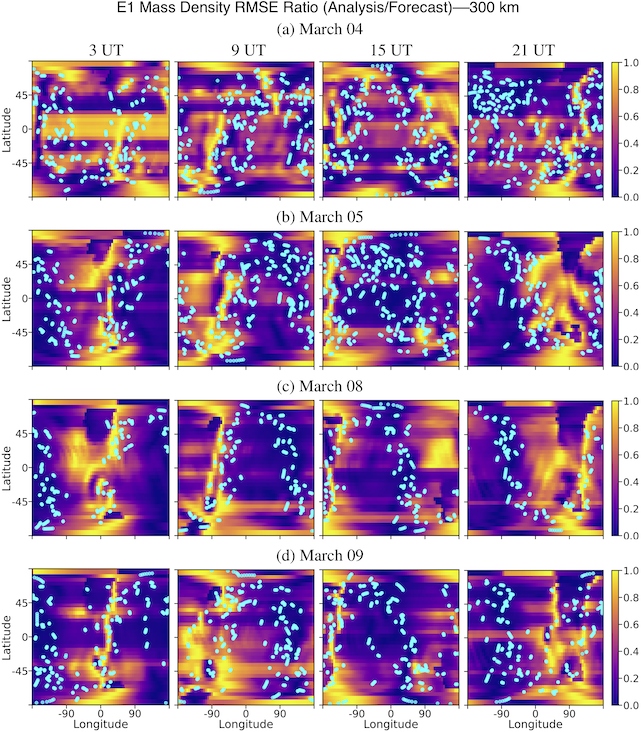}
\caption{Same as Figure~\ref{MarNEratio300} except for $R_{\textrm rmse}$ of mass density $\rho$ in E1.\label{MarDNratio300}}
\end{figure}

\begin{figure}
\centering
\includegraphics{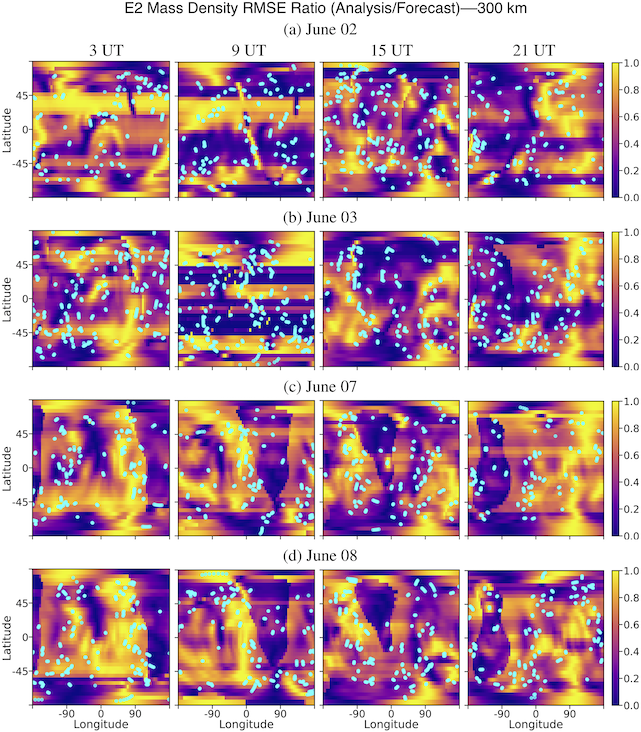}
\caption{Same as Figure~\ref{MarDNratio300} except for E2.\label{JunDNratio300}}
\end{figure}
Figures~\ref{MarDNratio300} and \ref{JunDNratio300} present the $R_{\textrm rmse}$($\rho$) at 300-km altitude for the experiments E1 and E2, respectively. Similar to Figures~\ref{MarNEratio300} and \ref{JunNEratio300}, the results for $R_{\textrm rmse}$($\rho$) also exhibit a correlation with assimilated COSMIC-$Ne$ profiles. The presence of purple patches in Figure~\ref{MarDNratio300} indicates that the assimilation impact in $R_{\textrm rmse}$($\rho$) is more localised compared to that of $R_{\textrm rmse}$($Ne$) for E1 in Figure~\ref{MarNEratio300}. Two hours into the assimilation (Figure~\ref{MarDNratio300}a [3~UT]), the improvement around the equator is significantly poor even with some COSMIC-$Ne$ profiles in the region. The situation is improved in the next snapshot in Figure~\ref{MarDNratio300}a [9~UT]. The analysis state is also degraded mainly in the southern latitudes between the longitudes 0 and 180{\textdegree}W at 9~UT on 9 March 2008, which is the storm day (see Figure~\ref{solgeoMar_Jun}a). In general, the $R_{\textrm rmse}$ is higher in areas with less-to-no observations in Figure~\ref{MarDNratio300}.

In Figure~\ref{JunDNratio300}a [21~UT], the longitudes to the west of 0{\textdegree} show more improvement than to the east. More COSMIC-$Ne$ profiles are also present in the longitudes to the west of 0{\textdegree} than to the east. Figure~\ref{JunDNratio300}d [21~UT] shows improved areas on the east mostly around observation locations. The $R_{\textrm rmse}$ in Figures~\ref{JunDNratio300}c--\ref{JunDNratio300}d compared to that of Figures~\ref{MarDNratio300}c--\ref{MarDNratio300}d indicates that the impact of assimilating COSMIC-$Ne$ on mass density during high solar activity with enhanced geomagnetic activity (7, 8 June 2014) is rather marginal. The differences in panels c and d in both Figures~\ref{MarDNratio300} and \ref{JunDNratio300} could also be due to the relatively less number of assimilated observations in the latter.
\subsection{Data-Guided Thermosphere Compared to Satellite Data}\label{sec_KFvaldtn}
\begin{figure}
\centering
\includegraphics[width=0.8\columnwidth]{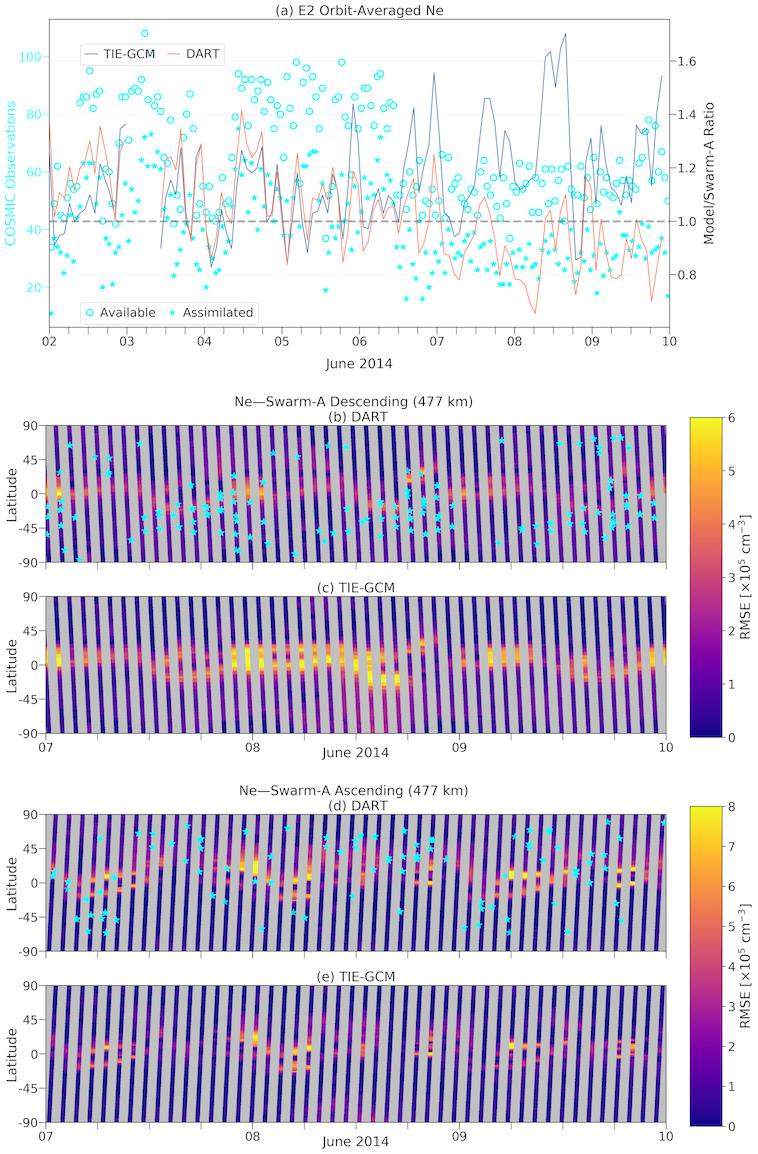}
\caption{(a-right) Ratio of Swarm-A orbit-averaged $Ne$ with TIE-GCM (blue) and E2 analysis mean $\boldsymbol{x}^{a}$ (DART; orange). The dashed line represents the ideal ratio. (a-left) The number of available/assimilated COSMIC-$Ne$ data between 450- and 550-km altitudes. The RMSE of E2-$\boldsymbol{x}^{a}$ and TIE-GCM with respect to the in-situ Swarm-A $Ne$ along the (b--c) descending and (d--e) ascending orbits. (b, d) The blue stars indicate the locations of assimilated COSMIC-$Ne$ in the vicinity of Swarm-A.\label{SwAratio_AcnDcn}}
\end{figure}

\begin{figure}
\centering
\includegraphics[width=0.8\columnwidth]{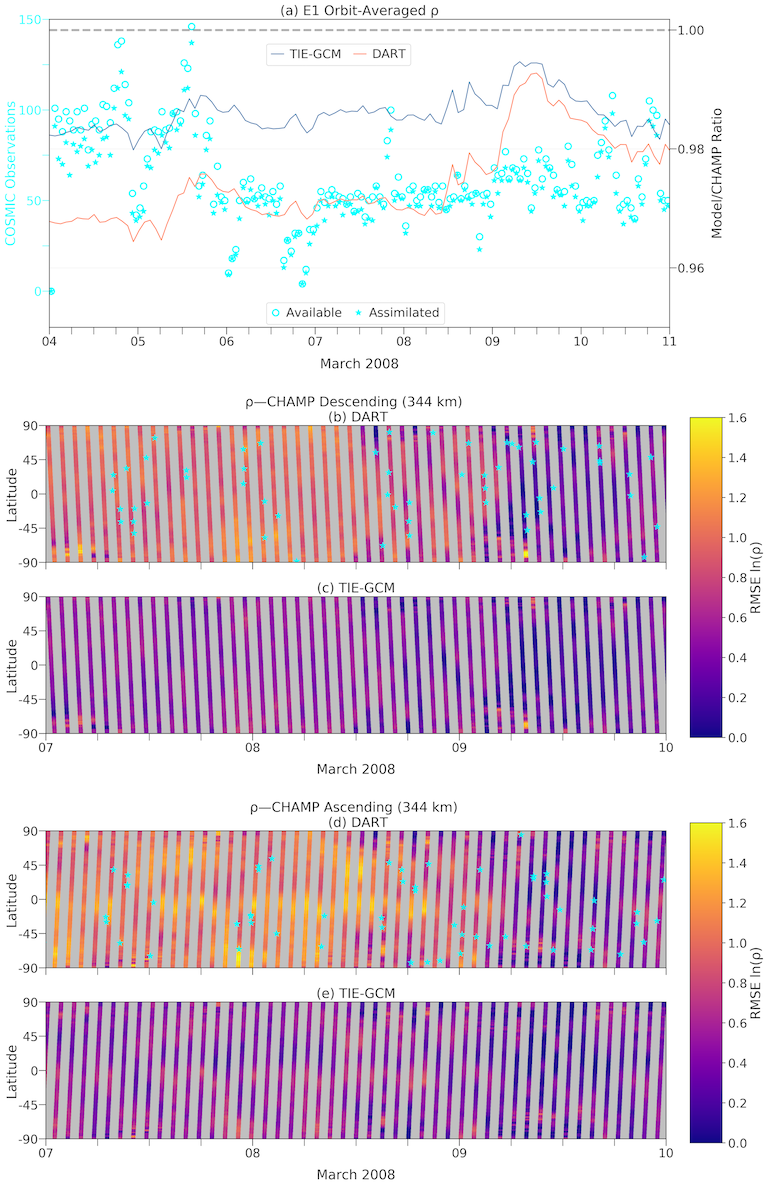}
\caption{Same as Figure~\ref{SwAratio_AcnDcn} except for accelerometer-derived mass density $\rho$ from CHAMP. Here RMSE ln($\rho$) of 1.6 is approximately 4.2$\times$10$^{-15}$~g$\cdot$cm$^{-3}$ in RMSE $\rho$. DART is $\boldsymbol{x}^{a}$ from E1. The number of available/assimilated COSMIC-$Ne$ data are between 250- and 350-km altitudes.\label{CHratio_AcnDcn}} 
\end{figure}

\begin{figure}
\centering
\includegraphics[width=0.8\columnwidth]{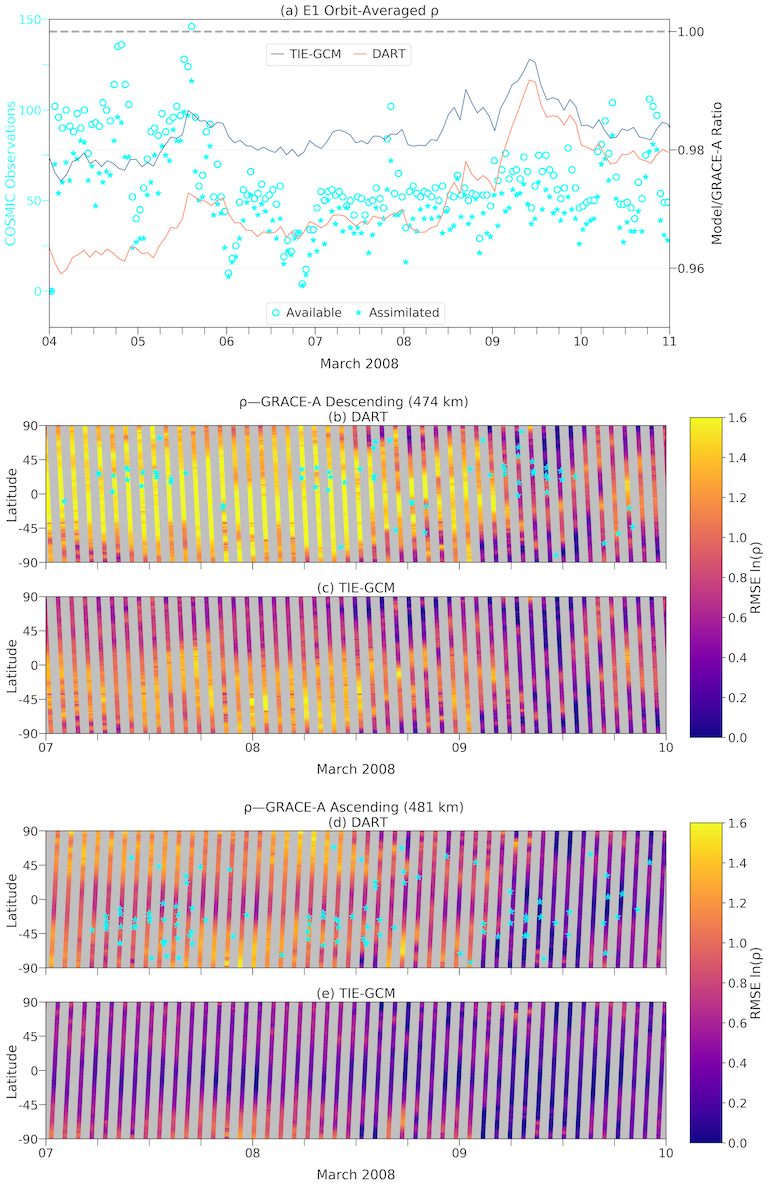}
\caption{Same as Figure~\ref{CHratio_AcnDcn} except for GRACE-A. Here RMSE ln($\rho$) of 1.6 is approximately 2.5$\times$10$^{-16}$~g$\cdot$cm$^{-3}$ in RMSE $\rho$. The number of available/assimilated COSMIC-$Ne$ data are between 450- and 550-km altitudes.\label{GrAratio_AcnDcn}}
\end{figure}

\begin{figure}
\centering
\includegraphics[width=0.8\columnwidth]{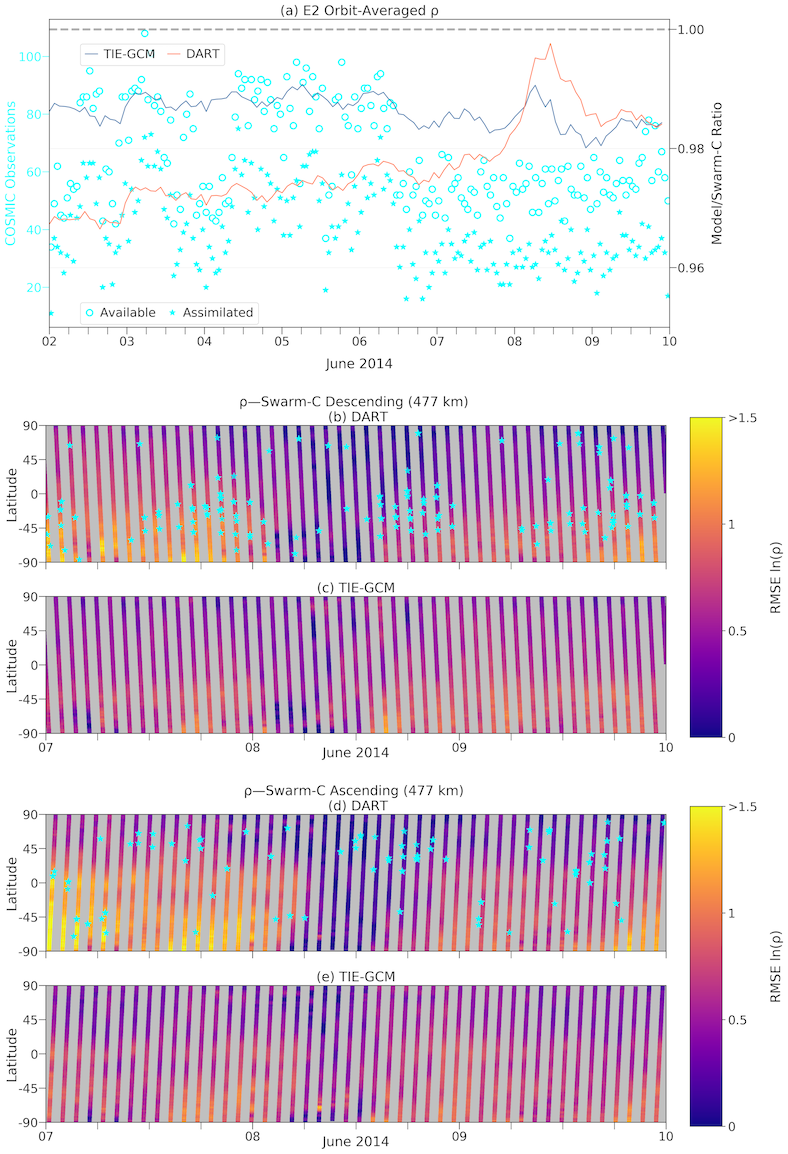}
\caption{Same as Figure~\ref{GrAratio_AcnDcn} except for Swarm-C in 2014. Here RMSE ln($\rho$) of 1.5 is approximately 6$\times$10$^{-16}$~g$\cdot$cm$^{-3}$ in RMSE $\rho$. DART is $\boldsymbol{x}^{a}$ from E2.\label{SwCratio_AcnDcn}}
\end{figure}

In this section, the COSMIC-$Ne$ assimilated analysis state from the two experiments E1 and E2 is interrogated with multiple independent sets of data. Here the analysis state is the above-mentioned mean of the updated/posterior ensemble $\boldsymbol{x}^{a}$ and is labelled as ``DART'' in Figures~\ref{SwAratio_AcnDcn}--\ref{SwCratio_AcnDcn}.

Figure~\ref{SwAratio_AcnDcn} compares $Ne$ measured aboard the Swarm-A satellite with TIE-GCM (aforementioned control state $\boldsymbol{x}^{c}$), and COSMIC-$Ne$-guided E2-$\boldsymbol{x}^{a}$ (DART). Figure~\ref{SwAratio_AcnDcn}a~[right] presents the TIE-GCM/Swarm-A (blue) and DART/Swarm-A (orange) orbit-averaged $Ne$ ratios. The line breaks correspond to the data-loss periods. Figure~\ref{SwAratio_AcnDcn}a~[left] shows the number of available (circle) and assimilated (star) COSMIC-$Ne$ profiles between the altitudes of 450 and 550 km.

In Figure~\ref{SwAratio_AcnDcn}a~[right], DART mostly overestimates the orbit-averaged $Ne$ compared to TIE-GCM for the first four days. While on average the number of assimilated observations is around 40 in Figure~\ref{SwAratio_AcnDcn}a~[left], one can discern that the ratio of observations not assimilated during the first four days is higher than that of the last four days. The drop in DART/Swarm-A ratio in Figure~\ref{SwAratio_AcnDcn}a~[right] seems to correspond with the drop in $Dst$ from late 7 June to 8 June 2014 in Figure~\ref{solgeoMar_Jun}b. In general, from 7 June 2014 onwards, DART performs better than TIE-GCM.

The panels b to d of Figure~\ref{SwAratio_AcnDcn} present the RMSE with respect to the in-situ Swarm-A $Ne$ during this better-performing period (7--10 June 2014) with each orbit separated into descending and ascending segments. The blue stars in panels b and d of Figure~\ref{SwAratio_AcnDcn} indicate the location of the assimilated COSMIC-$Ne$ observations in the vicinity of the satellite. The observations along the path of the satellite within $\pm$10{\textdegree} longitudinally, $\pm$10~min temporally, and $\pm$50~km altitudinally are considered to be in the vicinity of the satellite.

From the perspective of descending orbits in Figures~\ref{SwAratio_AcnDcn}b and \ref{SwAratio_AcnDcn}c, DART performs significantly better than TIE-GCM around the equator and low latitudes. The majority of the COSMIC-$Ne$ observations along the descending orbits are clustered across the equator and middle southern latitudes in the second half of each day. TIE-GCM's performance in the Southern Hemisphere is also degraded from approximately 12 to 18~UT on 8 June 2014 but DART's performance seems to be unaffected. A geomagnetic storm with a $Kp$ of 6 occurred on 8 June. The RMSE results presented in Figures~\ref{SwAratio_AcnDcn}d--\ref{SwAratio_AcnDcn}e show that the performance of the two models is comparable along the ascending orbits. In contrast to the pattern in Figure~\ref{SwAratio_AcnDcn}b, the observations in Figure~\ref{SwAratio_AcnDcn}d are mostly spread across the Northern Hemisphere and are relatively small in number. TIE-GCM also performs significantly better around the low latitudes in the ascending orbits compared to its performance in the region along the descending orbits. DART's performance is impressive given that it had no access to the exact GPIs that were available to TIE-GCM.

The three comparisons shown in Figures~\ref{CHratio_AcnDcn}, \ref{GrAratio_AcnDcn}, and \ref{SwCratio_AcnDcn} are analogous to Figure~\ref{SwAratio_AcnDcn} but for the accelerometer-derived mass densities from CHAMP, GRACE-A, and Swarm-C satellites, respectively. In these figures, the natural logarithm of mass density ln($\rho$) is used following, for example, \citet{Bruinsma_2018}, \citet{2002Picone}, and \citet{Sutton_tiegcm_assim}.  $\rho$ is given in units of g$\cdot$cm$^{-3}$. The label ``DART'' in Figures~\ref{CHratio_AcnDcn} and \ref{GrAratio_AcnDcn} correspond to $\boldsymbol{x}^{a}$ from E1, and likewise $\boldsymbol{x}^{a}$ from E2 in Figure~\ref{SwCratio_AcnDcn}. The number of available/assimilated COSMIC-$Ne$ profiles within $\pm$50~km of 300, 500, and 500~km altitudes are plotted in Figures~\ref{CHratio_AcnDcn}, \ref{GrAratio_AcnDcn}, and \ref{SwCratio_AcnDcn}, respectively. The available versus assimilated statistics reveal that the ratio of discarded observations is significantly lower during solar minimum (Figures~\ref{CHratio_AcnDcn}a~[left] and \ref{GrAratio_AcnDcn}a~[left]) compared to solar maximum (Figure~\ref{SwCratio_AcnDcn}a~[left]).

The increase in DART/CHAMP ratio in Figure~\ref{CHratio_AcnDcn}a~[right] during 5 March 2008 correlates with the increase in number of assimilated observations on the day. As the corresponding trend in TIE-GCM/CHAMP ratio is also very similar, this may not be a strong indicator of COSMIC-$Ne$-guided improvement but an enhancement of $\rho$ in data. In Figure~\ref{CHratio_AcnDcn}a~[right], DART has gradually matched the performance of TIE-GCM.

The ascending-descending decomposition along the CHAMP orbit provided in panels b to e of Figure~\ref{CHratio_AcnDcn} display the areas where the EAKF-based assimilation has minimised the RMSE of ln($\rho$). While DART's overall RMSE results are higher than that of TIE-GCM for the most of 7--10 June 2014, it is clear that in most areas where observations are assimilated (blue stars), the RMSE values have improved.

Figure~\ref{GrAratio_AcnDcn} provides validation of TIE-GCM and assimilation results against GRACE-A mass densities near the model's upper boundary. The model/data ratios corresponding to GRACE-A in Figure~\ref{GrAratio_AcnDcn}a~[right] are similar to the performance at CHAMP altitudes in Figure~\ref{CHratio_AcnDcn}a~[right]. However, DART in Figure~\ref{GrAratio_AcnDcn}a performs poorer than DART in Figure~\ref{CHratio_AcnDcn}a during 4 March 2008. Interestingly, the number of assimilated observations on 4 March 2008 is less in Figure~\ref{GrAratio_AcnDcn}a~[left] than in Figure~\ref{CHratio_AcnDcn}a~[left].

The RMSE ln($\rho$) for DART around the equator along the ascending orbits in Figure~\ref{GrAratio_AcnDcn}d is significantly lower compared to that of the descending orbits in Figure~\ref{GrAratio_AcnDcn}b. Figures~\ref{GrAratio_AcnDcn}c and \ref{GrAratio_AcnDcn}e reveal that compared to GRACE-A, TIE-GCM's largest excursions are mostly along the descending orbits. Both DART and TIE-GCM perform relatively poor along descending orbits compared to the ascending orbits. These differences could be due to the systematic biases (e.g. associated with satellite local solar time) inherent to the underlying TIE-GCM as well as systematic biases between ascending and descending segments of the accelerometer-derived mass densities. Data assimilation alone is perhaps not sufficient to mitigate the effects from these biases.

Figure~\ref{SwCratio_AcnDcn} provides a validation of the model with respect to Swarm-C mass densities during solar maximum. The trends in model/data ratio of orbit-averaged $\rho$ from both DART and TIE-GCM during this period are similar to that of during the solar minimum. Unlike in Figure~\ref{GrAratio_AcnDcn}, the RMSE separated into descending and ascending orbits in Figure~\ref{SwCratio_AcnDcn} reveals that in general, the performance along both portions of the orbit is similar---with some minor differences---for both DART and TIE-GCM. The colour-scale in Figure~\ref{SwCratio_AcnDcn} is culled at RMSE ln($\rho$) of 1.5 (RMSE $\rho$ of about 6$\times$10$^{-16}$~g$\cdot$cm$^{-3}$) to avoid a few outliers from skewing the colour-scale. The outliers are about 10 epochs along the Swarm-C orbit that produced an RMSE ln($\rho$) of approximately 2.5 (RMSE $\rho$ of about 8$\times$10$^{-16}$~g$\cdot$cm$^{-3}$).
\subsection{Impact of Assimilating In-situ Temperature}\label{sec_TNassim}
\begin{figure}
\centering
\includegraphics[width=\linewidth]{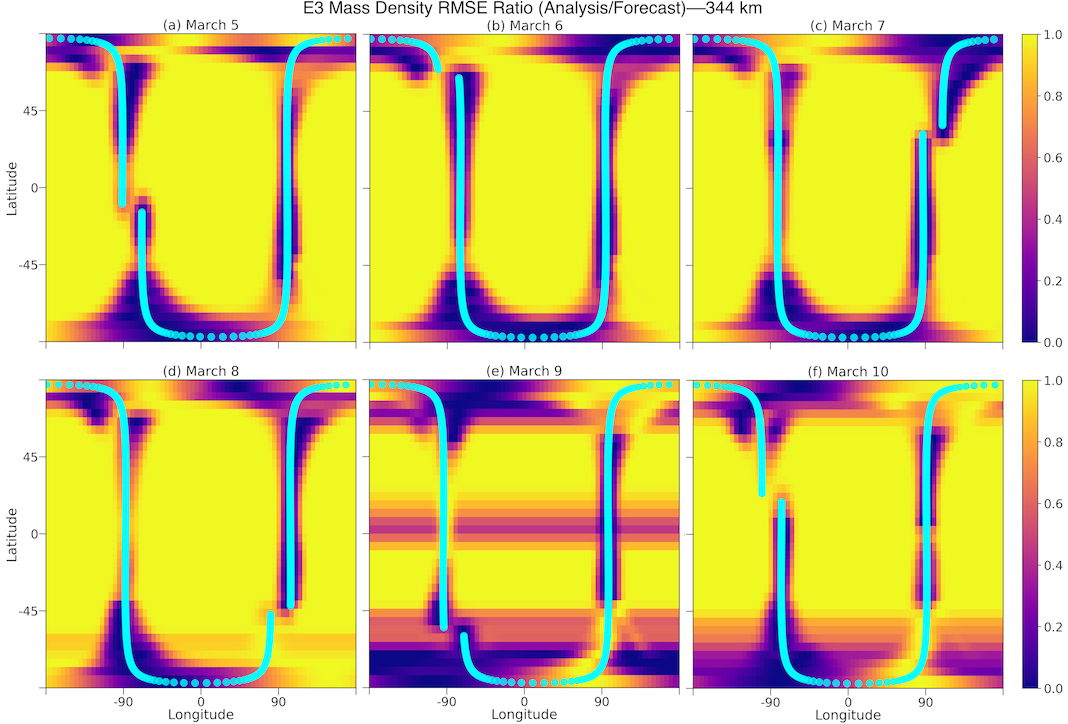}
\caption{The $R_{\textrm rmse}$($\rho$) for E3 at 344-km altitude. The blue dots indicate the latitude-longitude path of assimilated CHAMP-$Tn$ for each day within 90~min centred at 16~UT. The $R_{\textrm rmse}$($\rho$) are scaled from 0 to 1 where values close to 0 indicate that the analysis state $\boldsymbol{x}^{a}$ is closer to the ``truth''. \label{TNassim_rmse}}
\end{figure}

\begin{figure}
\centering
\includegraphics[width=0.8\columnwidth]{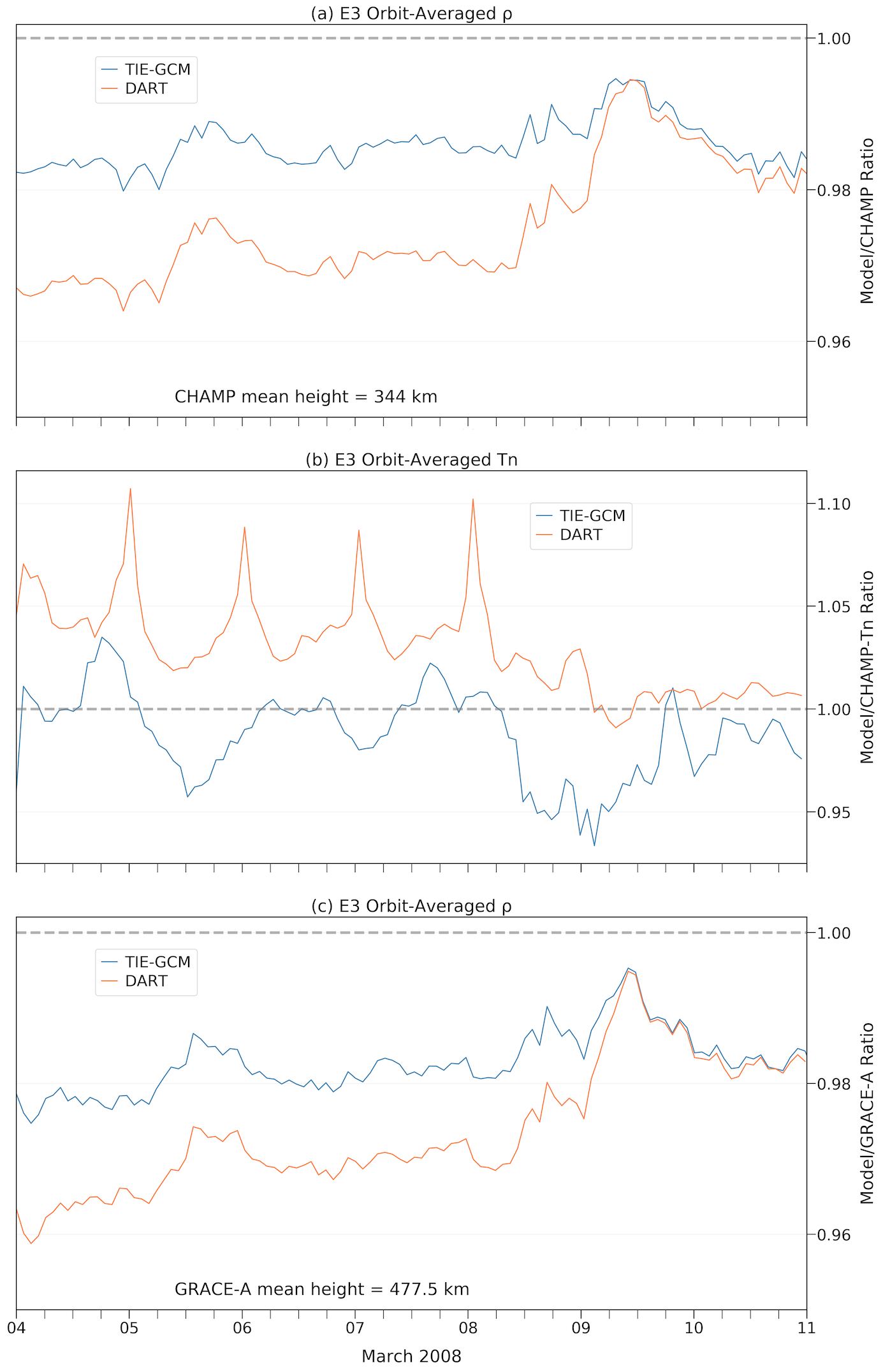}
\caption{Ratio of orbit-averaged (a) CHAMP mass density $\rho$, (b) CHAMP-$Tn$ temperature, and (c) GRACE-A $\rho$ with TIE-GCM (blue) and E3 analysis mean $\boldsymbol{x}^{a}$ (DART; orange). CHAMP-$Tn$ is the data assimilated in E3 to obtain the $\boldsymbol{x}^{a}$ state. The dashed line represents the ideal ratio.\label{TnAssim_CHGRratios}}
\end{figure}
Figure~\ref{TNassim_rmse} compares the RMSE ratios of mass density $R_{\textrm rmse}$($\rho$) (see Equation~(\ref{eqn_Rrmse})) for E3 at the average altitude of CHAMP (344~km). The figure shows the impact of the assimilation of CHAMP-$Tn$ data at 16~UT daily from 5 to 10 March 2008. In Figure~\ref{TNassim_rmse}, the $R_{\textrm rmse}$($\rho$) is scaled from 0 to 1, where 0 indicate that the analysis $\boldsymbol{x}^{a}$($\rho$) is statistically indistinguishable from the control state $\boldsymbol{x}^{c}$---the ``truth''. The blue dots follow the latitude-longitude path of the assimilated CHAMP-$Tn$ data points within 90~min centred at 16~UT along the CHAMP orbit. The geographic latitude-longitude resolution of the figure is 5{\textdegree}$\times$5{\textdegree}.

Figure~\ref{TNassim_rmse} shows that assimilation of in-situ CHAMP-$Tn$ has reduced the $R_{\textrm rmse}$ in the analysis state of mass density mostly along the CHAMP track. In other words, the impact is local unlike the global impact shown, for example, in Figure~\ref{MarDNratio300} from assimilating COSMIC-$Ne$. This behaviour is due to the fact that the assimilated CHAMP-$Tn$ data are spatially contained to the orbit. The horizontal localisation used in both Figures~\ref{MarDNratio300} and \ref{TNassim_rmse} (E1 and E3) is the same.

Figure~\ref{TnAssim_CHGRratios} compares the results from E3 to accelerometer-derived mass densities from CHAMP and GRACE-A, and temperature data from CHAMP-$Tn$---the same data set assimilated in E3. As described in Section~\ref{sec_KFvaldtn}, the model/data ratios presented here are also computed using the natural logarithm of mass densities. The label ``DART'' in Figure~\ref{TnAssim_CHGRratios} corresponds to $\boldsymbol{x}^{a}$ from E3. TIE-GCM in Figure~\ref{TnAssim_CHGRratios} is the aforementioned control state $\boldsymbol{x}^{c}$ driven with observed GPIs.

Figure~\ref{TnAssim_CHGRratios}a shows that DART/CHAMP ratio converges to the respective TIE-GCM ratio. The TIE-GCM blue lines in Figures~\ref{CHratio_AcnDcn}a and \ref{TnAssim_CHGRratios}a are the same. Although the DART orange lines in the two figures appear similar, the DART results corresponding to E3 (Figure~\ref{TnAssim_CHGRratios}a) slightly outperform that of E1 during 9--10 March 2008 (Figure~\ref{CHratio_AcnDcn}a~[right]). Likewise, the DART in Figure~\ref{TnAssim_CHGRratios}c outperforms the DART in Figure~\ref{GrAratio_AcnDcn}a~[right]. The DART in Figure~\ref{TnAssim_CHGRratios}c shows that the continuous assimilation of the temperature data at CHAMP altitudes (mean orbital height of 344~km) is capable of influencing the mass density specification even at the GRACE-A altitudes (mean orbital height of 477.5~km).

Figure~\ref{TnAssim_CHGRratios}b, similar to the comparison in Figure~\ref{SwAratio_AcnDcn}a~[right], shows the impact of the assimilation on the same type of thermospheric parameter as the assimilated type. DART in Figure~\ref{TnAssim_CHGRratios}b demonstrates that EAKF makes large adjustments to the model temperature state during the first four days and then gradually settle at near-CHAMP-$Tn$ values. In other words, the ensemble mean-state of temperature in E3 displays a significant daily variation until 8 March 2008 compared to the last two days of the experiment. TIE-GCM in Figure~\ref{TnAssim_CHGRratios}b mostly underestimates the orbit-averaged temperature---the lines above the ideal line at 1.00 indicates the brief periods of overestimate. Daily variation of TIE-GCM is also relatively low on 10 March 2008.
\section{Discussion}{\label{discussion}}
This section discusses the results presented above in Sections~\ref{sec_Neassim}, \ref{sec_KFvaldtn}, and \ref{sec_TNassim}.

Assimilation of COSMIC-$Ne$ profiles decreases the $R_{\textrm rmse}$ for $Ne$ and $\rho$ more significantly and broadly during the solar minimum (E1) than the solar maximum (E2). The latitude-longitude maps of the $R_{\textrm rmse}$ for both E1 and E2 (Figures~\ref{MarNEratio300}--\ref{JunDNratio300}) reveal the differential model bias during the two vastly different solar activity periods. Although the compared periods belong to two different seasons (March-equinox and June-solstice months), the model bias due to the season in TIE-GCM is less compared to solar activity---$F_{10.7}$ \citep[e.g.][]{Emmert2014_msisTIEGCM}. Still, some seasonal bias may be present in these $R_{\textrm rmse}$ maps. The results presented in Figures~\ref{MarDNratio300} and \ref{JunDNratio300} show that the TIE-GCM performs better during the solar minimum than the solar maximum. \citet{Elvidge2016_MME} also reported that the mass density forecast skill of TIE-GCM is significantly better during the solar minimum than the solar maximum.

The $R_{\textrm rmse}$ maps present further insights into model bias in the two experiments E1 and E2. For example, the persistent quasi-terminator feature near 0{\textdegree} longitude in Figure~\ref{MarDNratio300}~[3~UT] is a clear indication of systematic bias. These quasi-terminator features are also present in $R_{\textrm rmse}$ maps for solar maximum---faintly in Figure~\ref{JunNEratio300} and slightly more enhanced in Figure~\ref{JunDNratio300}. Typically the ageostrophic/horizontal winds at these high altitudes also converge around the 0{\textdegree} longitude (at 3~UT) region primarily as a result of ion drag \citep{Hsu_EIA_2016}. \citet{Hsu_EIA_2016}, using TIE-GCM, shows that the temperature troughs, which occur around the wind-convergent region, is larger in amplitude during solar-maximum compared to the solar minimum. In TIE-GCM, the ion drag force is mainly controlled by $Ne$ (due to the ion-electron quasi-neutrality in the ionosphere). As the mass density is directly proportional to temperature, it is hypothesised that these features along the quasi-terminator region are more pronounced in the $\rho$ state than in $Ne$ state itself due to the additional temperature variations introduced by the assimilated $Ne$. This further emphasises the important role of ion-neutral coupling and criticality of correct specification of plasma-neutral interactions in models as relatively small errors in temperature can lead to relatively large errors in mass density \citep{Hsu_EIA_2016}. Likewise, to mitigate these systematic biases, calculation of ion/viscous drag forces may also require some adjustment. Reporting on assimilating $Ne$ into TIE-GCM, \citet{Matsuo_EnKF_2013} also alluded to (but did not investigate) the bias around thermospheric and ionospheric features with sharp spatial gradients such as the day-night terminator. The increase in $R_{\textrm rmse}$ along the quasi-terminator in Figure~\ref{MarDNratio300}~[3~UT] is perhaps one such example.

The assimilation window in E1 and E2 is 1~hr centred at the assimilation time---resulting in essentially assimilating data from the ``future''. In order to demonstrate this effect, COSMIC-$Ne$ profiles 30~min into the future are also included in the data locations shown in Figures~\ref{MarNEratio300}--\ref{JunDNratio300}. Moreover, considering the memory of the thermosphere of the adjusted state, COSMIC-$Ne$ profiles since 2.5~hr prior to the beginning of the assimilation window are also shown. Previous similar studies \citep[e.g.][]{LeeMatsuo_2012_cosmic,Hsu2014} showed that the adjustment to the $Ne$ state via data assimilation disappears shortly on the order of hours and the ionosphere has a tendency to relax toward climatology in the model. \citet{Jee_2007}, using the thermosphere-ionosphere nested grid model, showed that the time it takes for the ionosphere to recover from an altered $Ne$ state is about 2--3 hr, where recovery is measured by the interval of time for difference between the original and altered states to decrease by a factor of $e$. \citet{Jee_2007}'s results also show that this recovery time is highly dependent on latitude and local time. The $R_{\textrm rmse}$ results in the above experiments also displayed specific regions where overlaid COSMIC-$Ne$ profiles did not seem to reduce the model error at certain times.  In other words, in those regions in Figures~\ref{MarNEratio300}--\ref{JunDNratio300}, the impact from the COSMIC-$Ne$ profiles on the model state may have already disappeared at the chosen snapshots.

Apart from assessing the impact of assimilating COSMIC-$Ne$ profiles into TIE-GCM, this study also measures the accuracy of the analysis state against independent satellite observations. DART in Figure~\ref{SwAratio_AcnDcn}a~[right] is slightly worse in estimating $Ne$ along Swarm-A than TIE-GCM at the start of the experiment but quickly matches up to TIE-GCM. More importantly, it does not deteriorate much in performance during storm time---8 June 2014. TIE-GCM, on the other hand, significantly departs from the ideal ratio during 6--8 June 2014. The time series also shows that about half of the available COSMIC-$Ne$ observations are discarded---this is due to the outlier threshold setting. The observations that are not within three standard deviations with the prior ensemble estimate are rejected. The EAKF is not expected to align the initial ensemble with the \textit{attractor} (estimate of the ``true'' state of the system generated from observations in a given assimilation window) instantaneously. The EAKF is designed to gradually coerce the initial ensemble to be consistent with the observations. In order to lower the observation rejection ratio, the observation error variance could be increased---at the cost of observation gain. In other words, the observation error tested in E1 and E2 could be increased from 15\% to a higher value to increase the number of assimilated observations. Although higher observation error may lower the observation gain, the increase in the number of observations may improve the overall result.

The comparison with independent satellite data also gives an indication about the limitations of assimilating COSMIC-$Ne$ profiles in an operational setting. While the number of profiles to assimilate can be controlled in experiments with synthetic observations \citep[e.g.][]{Matsuo_EnKF_2013,Hsu2014}, the availability of bona-fide COSMIC-$Ne$ profiles is dependent on multiple factors such as the number of transmitters in view, quality of the receiving radio signals, etc. As illustrated through isolating COSMIC-$Ne$ profiles along the ascending and descending orbits (see Figures~\ref{SwAratio_AcnDcn}--\ref{SwCratio_AcnDcn}), the number of observations that fall within the respective satellite's vicinity per orbit is sparse. This shows that in practice, for example, to predict mass density along a given orbit, the current spatial and local time coverage of COSMIC-$Ne$ profiles is rather poor. Future missions to add more capabilities and satellites to the COSMIC constellation may help improve this spatial and local time coverage. A COSMIC follow-on mission, FORMOSAT-7/COSMIC-2 is expected to be operational in 2019 with six satellites orbiting the equator at an inclination of 23{\textdegree} \citep[e.g.][]{Yue_cosmic_2014}.

Figures~\ref{SwAratio_AcnDcn}c and \ref{SwAratio_AcnDcn}e help precisely identify the segments of the orbit where TIE-GCM consistently registers larger errors. It is found that assimilation of COSMIC-$Ne$ has improved the model $Ne$ state around the equator in the descending segments (Figure~\ref{SwAratio_AcnDcn}b) compared to that of ascending segments (Figure~\ref{SwAratio_AcnDcn}d). This could be due to the availability of more COSMIC-$Ne$ profiles near the equator in Figure~\ref{SwAratio_AcnDcn}b as well as other systematic biases (e.g. associated with satellite local solar time, and biases between ascending and descending segments of the accelerometer-derived data).

The DART's improvement in estimating $\rho$ is gradual. The model/data ratio for orbit-averaged $\rho$ in panel a in Figures~\ref{CHratio_AcnDcn}--\ref{SwCratio_AcnDcn} does not seem to vary as much as the orbit-averaged $Ne$ in Figure~\ref{SwAratio_AcnDcn}. While DART's overall RMSE results are larger than that of TIE-GCM for the most of 7--10 June 2014 in Figure~\ref{SwCratio_AcnDcn}, it is clear that in most areas where observations are assimilated (blue stars), the RMSE values have improved. In general, DART shows some promise in improving the specification of $\rho$ via the assimilation of COSMIC-$Ne$.

The geometric heights in TIE-GCM is calculated using an empirical formulation relating spatially varying gravity with temperature and composition \citep{Qian201473}. The model-data comparisons presented here (e.g. Figures~\ref{SwAratio_AcnDcn}--\ref{SwCratio_AcnDcn} and Figure~\ref{TnAssim_CHGRratios}) are produced by interpolating DART and TIE-GCM along the satellite orbit contingent on the geometric heights mentioned above. As the native vertical coordinate system in TIE-GCM is based on atmospheric pressure, a similar height conversion is done when assimilating COSMIC-$Ne$ profiles into the model as COSMIC-$Ne$ profiles do not include measured atmospheric pressure. These height uncertainties are further accentuated at higher altitudes as the conversion to geometric height expands the vertical resolution at higher altitudes between consecutive pressure layers. In other words, near the lower boundary, pressure levels converted to geometric height have a resolution of about 3~km, but pressure levels around 300-km altitude typically have a resolution greater than 30~km. The uncertainties introduced by this back-and-forth height conversions require further investigation.

The results in Figure~\ref{TNassim_rmse} bear some resemblance to the results in Figure~2 of \citet{Matsuo_EnKF_2013}, which shows the impact of assimilating mass density from CHAMP. In Figure~\ref{TNassim_rmse}, the results from experiment E3 show that assimilation of in-situ CHAMP-$Tn$ into TIE-GCM improves the mass density specification mostly along the CHAMP satellite track. The overall impact of the assimilation is more significant in the middle to high latitudes than in low latitudes. 

The spread of low $R_{\textrm rmse}$($\rho$) in the middle to high latitudes in Figure~\ref{TNassim_rmse} may have some contribution from the fact that CHAMP's dwelling time over the high latitudes is greater than that of over the equator (hence more measurement epochs). This spread is also due to the fact that the effective area in the polar regions is smaller compared to the equator---longitudes converging at the poles. The band across the equator in Figure~\ref{TNassim_rmse}e corresponds to the geomagnetic storm on 9 March 2008 as illustrated in Figure~\ref{solgeoMar_Jun}a. It is unlikely that assimilated CHAMP-$Tn$ effected such change in mass density encompassing the equatorial region.

As mentioned above, CHAMP-$Tn$ data are not bona-fide in-situ measurements but empirically derived in the recalibration of CHAMP $\rho$ data by \citet{Mehta_2017}. The results in Figure~\ref{TnAssim_CHGRratios} suggest that assimilating such along-orbit data from a particular satellite may prove even more useful in forecasting applications involving the same satellite than assimilating sparse global observations. Temperature is a critical parameter in the specification of mass density in the thermosphere. Figure~\ref{TnAssim_CHGRratios}c shows that altering the temperature state in a limited region (e.g. along the CHAMP orbit) has the potential to effect change even at a higher altitude.
\section{Summary and Conclusions}{\label{conclusion}}
The paper presented two experiments of assimilating electron densities and one experiment of assimilating temperature into TIE-GCM using the EAKF technique. A study of the EAKF/TIE-GCM framework for the thermosphere not only helps to gauge the accuracy of the assimilation, to explain the inherent model bias, and to understand the limitations of the framework, but it also establishes EAKF as a viable technique in the presence of realistic data assimilation scenarios to forecast the highly dynamical thermosphere.

The results from perfect model scenarios showed that model state is changed and, more often than not, improved with data assimilation---reduced $R_{\textrm rmse}$ ratios. The resulting analysis states are validated against independent data from multiple satellites including Swarm-A, Swarm-C, CHAMP, and GRACE-A. The validation results showed that the COSMIC-$Ne$-guided thermosphere state does not outperform the GPI-guided TIE-GCM along the considered orbits. This may be due to the limited number of bona-fide COSMIC-$Ne$ profiles available for the thermosphere specification tasks in the experiments. 

The experiments E1 and E2 indicated that using COSMIC-$Ne$ profiles in an operational setting is challenging and that the area and local time coverages of the profiles are perhaps too sparse to be used in, for example, the applications of orbit prediction. The CHAMP-$Tn$-guided experiment showed more promise over COSMIC-$Ne$ in terms of estimating mass density along the orbits of both CHAMP and GRACE-A satellites.

The results in Figures~\ref{MarNEratio300}--\ref{JunDNratio300} showed that the improvement gained in the overall forecasted thermosphere state is better during solar minimum compared to that of solar maximum. These results also provided insights into the biases inherent in TIE-GCM---particularly along thermospheric features with sharp spatial gradients. The systematic biases that above results highlighted could be an indication that the specification of plasma-neutral interactions in TIE-GCM needs further adjustments.

The experiments mainly focused on the assimilation accuracy during different solar activity periods. More work needs to be done to identify and improve model bias due to external forcing. Assimilation of other thermospheric data, for example, ground-based remote sensing measurements of thermospheric neutral winds and temperature could also help unravel some of the difficulties associated with forecasting thermospheric mass density. The new GOLD (Global-scale Observations of the Limb and Disk) mission will also provide space-based measurements of thermospheric composition and temperature data to conduct these assimilation experiments.

\section*{\small{Acknowledgements}}
\small{This research was partially supported by a research scholarship awarded to T.~Kodikara by the Cooperative Research Centre for Space Environment Management, SERC Limited, through the Australian Government's Cooperative Research Centre Programme. This research was also partially supported by the National Natural Science Foundation of China (project ID: 41730109) and the Jiangsu dual creative talents and teams programme projects awarded in 2017. This research was undertaken with the assistance of resources from the National Computational Infrastructure (NCI), which is supported by the Australian Government.

The measurements $Kp$, $ap$, $Dst$ and $F_{10.7}$ are obtained from the OMNI database: \url{omniweb.gsfc.nasa.gov}. The DART assimilation tools used in this work are available for free at \url{http://www.image.ucar.edu/DAReS/DART}. The TIE-GCM is available for free at \url{www.hao.ucar.edu/modeling/tgcm}. The electron density profiles are obtained from the COSMIC Data Analysis and Archive Center at the University Corporation for Atmospheric Research (\url{cosmic.ucar.edu}). Swarm data can be accessed from ESA at \url{https://earth.esa.int/web/guest/swarm/data-access}. Piyush Mehta is thanked for sharing the CHAMP and GRACE-A data via \url{http://tinyurl.com/densitysets}. Alex Chartier is thanked for his insightful discussions during the early stages of this study. Tim Hoar is thanked for his valuable advice on data assimilation using DART during various stages of the study. Brett Carter, Robert Norman, and Donald Grant are thanked for their constructive comments and helpful suggestions.}
\bibliographystyle{apacite}

\end{document}